%% file: main.tex
\documentclass[conference]{IEEEtran}
\usepackage{cite}
\usepackage{amsmath,amssymb,amsfonts}
\usepackage{algorithm}
\usepackage{algorithmic}
\usepackage{graphicx}
\usepackage{textcomp}
\usepackage{xcolor}
\usepackage[hyphens]{url}
\usepackage{fancyhdr}
\usepackage{hyperref}
\usepackage{array}
\usepackage{tabularx}
\usepackage{bm}
\usepackage{booktabs}
\usepackage{authblk}
\usepackage[switch]{lineno}
\usepackage{multirow}
\usepackage{adjustbox}
\usepackage{float}
\usepackage{makecell, hhline}

\usepackage{graphicx}
\usepackage{multirow}
\usepackage{adjustbox}
\usepackage{amsmath}
\usepackage{amssymb}
\usepackage{booktabs}

\usepackage{cite}
\usepackage{amsmath,amssymb,amsfonts}
\usepackage{algorithmic}
\usepackage{graphicx}
\usepackage{textcomp}
\usepackage{xcolor}
\usepackage{hyperref}
\usepackage[a4paper, total={184mm,239mm}]{geometry}

\def\BibTeX{{\rm B\kern-.05em{\sc i\kern-.025em b}\kern-.08em
    T\kern-.1667em\lower.7ex\hbox{E}\kern-.125emX}}

\title{Hyperdimensional Intelligent Sensing for Efficient Real-Time Audio Processing on Extreme Edge}

\author{Sanggeon Yun$^1$, Ryozo Masukawa$^1$, Hanning Chen$^1$, SungHeon Jeong$^1$, Wenjun Huang$^1$\\Arghavan Rezvani$^1$, Minhyoung Na$^2$, Yoshiki Yamaguchi$^3$ and Mohsen Imani$^1{}^{\star}$ \\ $^1$University of California, Irvine, CA, USA\\ $^2$Kookmin University, Seoul, Republic of Korea\\ $^3$Shibaura Institute of Technology, Saitama, Japan\\
$^{\star}$Corresponding Author: m.imani@uci.edu \vspace{-3mm}}

\begin{document}
\maketitle
\thispagestyle{plain}
\pagestyle{plain}

\begin{abstract}
    \input{Chapters/0_Abstract}
\end{abstract}


\section{Introduction}\label{sec:introduction}

\input{Chapters/1_Introduction}

\section{Background}\label{sec:background}

\input{Chapters/2_Background}

\section{Intelligent Sensing Model Design}\label{sec:intelligent_sensing_model_design}

\input{Chapters/3_Model_Design}

\section{Experiments}\label{sec:experiments}

\input{Chapters/4_Experiments}

\section{Conclusions}\label{sec:conclusions}

\input{Chapters/5_Conclusions}

\section*{Acknowledgements}
This work was supported in part by DARPA Young Faculty Award, National Science Foundation \#2127780, \#2319198,  \#2321840 and \#2312517, Semiconductor Research Corporation (SRC), Office of Naval Research, grants \#N00014-21-1-2225 and \#N00014-22-1-2067, the Air Force Office of Scientific Research under award \#FA9550-22-1-0253, and generous gifts from Xilinx and Cisco.

\bibliographystyle{IEEEtranS}
\bibliography{refs}

\end{document}

%% file: Chapters/0_Abstract.tex
The escalating challenges of managing vast sensor-generated data, particularly in audio applications, necessitate innovative solutions. Current systems face significant computational and storage demands, especially in real-time applications like gunshot detection systems (GSDS), and the proliferation of edge sensors exacerbates these issues. This paper proposes a groundbreaking approach with a near-sensor model tailored for intelligent audio-sensing frameworks. Utilizing a Fast Fourier Transform (FFT) module, convolutional neural network (CNN) layers, and HyperDimensional Computing (HDC), our model excels in low-energy, rapid inference, and online learning. It is highly adaptable for efficient ASIC design implementation, offering superior energy efficiency compared to conventional embedded CPUs or GPUs, and is compatible with the trend of shrinking microphone sensor sizes. Comprehensive evaluations at both software and hardware levels underscore the model's efficacy. Software assessments through detailed ROC curve analysis revealed a delicate balance between energy conservation and quality loss, achieving up to 82.1\% energy savings with only 1.39\% quality loss. Hardware evaluations highlight the model's commendable energy efficiency when implemented via ASIC design, especially with the Google Edge TPU, showcasing its superiority over prevalent embedded CPUs and GPUs.

%% file: Chapters/1_Introduction.tex
In today's systems, the challenge of handling the large amounts of data generated by sensors, particularly in the context of audio data, has become increasingly prevalent~\cite{zaman2023survey}. This surge in data volume requires significant computational resources for both processing and storage, at significant cost. This is particularly problematic in domains that require real-time responsiveness, such as gunshot detection systems (GSDS)~\cite{hansen2021gunshot}, chainsaw sound detection for forest protection~\cite{somwong2023acoustic}, and vehicle sound detection for real-time traffic planning~\cite{10078155}. In addition, the proliferation of edge sensors in today's landscape further exacerbates the cost scaling of these systems.

Nevertheless, in many of these systems, a substantial portion of the data generated by microphone sensors is irrelevant to the targeted detection scenarios~\cite{calhoun2021precision}. It is plausible to mitigate these escalating costs by transmitting only the pertinent audio data of interest, which is essential for predicting targeted detection scenarios, to cloud-based systems responsible for data storage and processing. One promising solution for achieving this efficiency is the adoption of near-sensor computing paradigms. By embedding lightweight, real-time audio detection machine learning models with low energy consumption capabilities at the sensor level, we can selectively forward valuable data to the cloud infrastructure.

Contemporary deep learning models, such as Deep Neural Networks (DNNs), have gained significant traction in addressing various real-world tasks. However, they have been critiqued for their substantial memory and energy requirements, particularly during the training and inference phases. Deploying such resource-intensive models directly on or near sensors poses challenges, resulting in a notable performance gap between deep learning algorithms and sensing components. Moreover, DNNs are less adaptable to online learning scenarios, a critical limitation in systems that must promptly adjust to real-time sensory input.

The aforementioned challenges serve as a strong motivation for the development of an intelligent, rapidly adaptable, and efficient framework for the representation and analysis of raw sensor data. To attain real-time performance, even with online learning capabilities, we propose a redesign of machine learning algorithms. This involves the fusion of a lightweight Convolutional Neural Network (CNN) with neurally-inspired HyperDimensional Computing (HDC). HDC is an alternative paradigm that emulates essential brain functions, prioritizing high efficiency and online learning capacity, while the CNN is harnessed for its robust feature extraction capabilities. HDC is rooted in the observation that the human brain excels in operating on high-dimensional data representations.

In this paper, we introduce a near-sensor model designed for an intelligent sensing framework tailored to audio detection tasks, aiming to address the challenges previously delineated. Our model incorporates a Fast Fourier Transform module, followed by a series of CNN layers and an HDC layer. Notably, it facilitates low-energy, rapid inference, and online learning, dynamically adapting to emerging data trends based on feedback from a heavyweight machine learning model in the cloud. The versatility of our model extends to its adaptability for implementation in ASIC design, offering superior energy efficiency compared to embedded CPUs or GPUs. Given the current trend of shrinking microphone sensor sizes, an ASIC implementation supporting compact chip dimensions ensures the viability of deploying our proposed near-sensor model across a spectrum of contemporary sensors.

Subsequently, we conduct a comprehensive evaluation of our near-sensor model at both the software and hardware levels. On the software front, we present a detailed ROC curve analysis, providing a nuanced understanding of how our framework balances energy conservation against potential quality loss. Additionally, we demonstrate the model's performance in detecting audio of interest, showcasing its comparative efficacy with a larger, more complex cloud-based model. Shifting the focus to the hardware dimension, we underscore the high energy efficiency of our model through ASIC design implementation, comparing favorably against commonly employed embedded CPUs and GPUs. This dual-tier evaluation illuminates the prowess of our near-sensor model, affirming its viability and efficiency across both software and hardware realms.

Our work is fundamentally novel and provides the following contributions:
\begin{itemize}
    \item To the best of our knowledge, we propose the very first novel framework for saving total energy consumption of the audio detection domain systems where the audio of interest is infrequent.
    \item We propose an efficient near-sensor model that is capable of detecting audio of interest while supporting online learning. We further show a thorough analysis of the model to identify the proper hyperparameter for balancing the trade-off relationship between energy saving and quality loss.
    \item We show its energy efficiency when we run our proposed model on ASIC chip compared to generally used embedded CPUs and GPUs.
\end{itemize}

%% file: Chapters/2_Background.tex
\begin{figure*}
  \centering
  \includegraphics[width=0.6\linewidth]{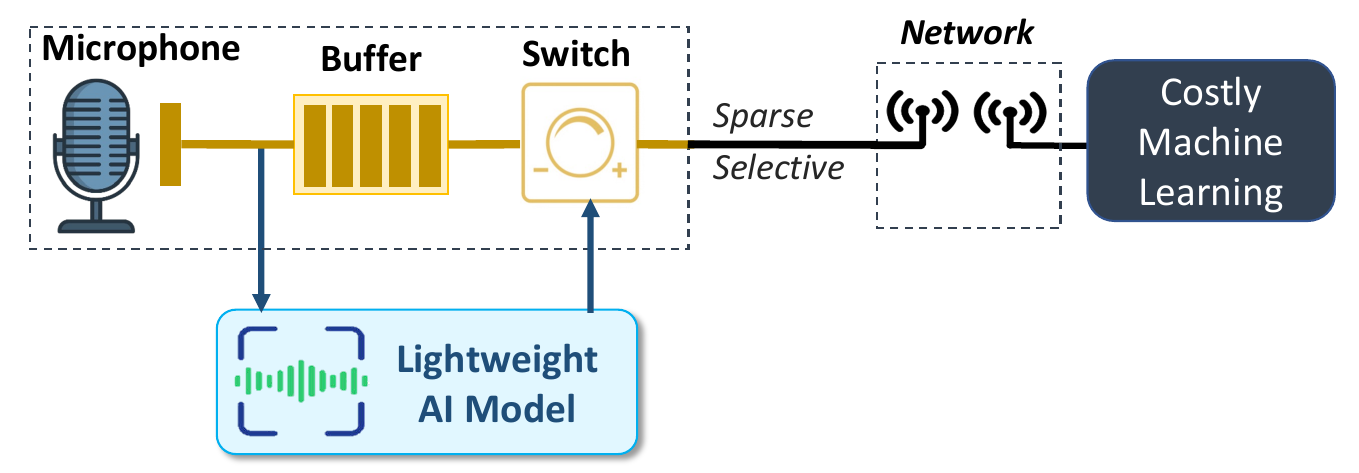}
  \caption{Overview of our Hyperdimensional Intelligent Sensing pipeline. The  ``sparse selective strategy'' is applied at the near-sensor stage, where only audio segments identified as audio-of-interest are transmitted to the cloud.}
  \label{Fig:overview_diagram_ours}
\end{figure*}

\subsection{Intelligent Sensing over Audio Data} 
With the continual advancements in sensor technology, there has been a parallel evolution in computational methods, aiming to extract meaningful insights from raw sensor data~\cite{ballard2021machine}. Notably, recent innovations in the form of in-sensor and near-sensor accelerators~\cite{li2021recent,angizi2022pisa,ma2022hogeye, sumbul2022system,zhou2020near, yun2024hypersense} represent a stride towards more efficient local processing. These accelerators integrate dedicated machine learning circuits into the sensing circuitry, enhancing the efficiency of data processing at the source. However, despite the success of these localized approaches, there is often a gap when it comes to achieving comprehensive system-level integration. In the specific context of audio data, recent research endeavors~\cite{Vajpayee_2023_WACV,10005289,10059143} have delved into efficient systems tailored to audio applications. These frameworks have tackled challenges ranging from memory reduction to devising low-power solutions. While these contributions showcase valuable progress, they may fall short in meeting the real-time demands of critical applications, such as acoustic gunshot detection~\cite{hansen2021gunshot}. The need for 24/7 monitoring and rapid response in scenarios like gunshot detection necessitates a paradigm shift towards real-time operation, a facet not thoroughly addressed by existing solutions.

To explicitly address these gaps, we note that existing methods either lack robust online learning capabilities at the edge or fail to drastically reduce energy consumption for large-scale deployments. Our proposed model directly tackles these shortcomings by introducing a near-sensor paradigm that can continuously adapt to new data while significantly cutting down on end-to-end system energy usage.

It is in this gap that our proposed near-sensor model stands as a pioneering framework. Going beyond existing approaches, our model is designed to minimize the overall energy consumption associated with audio data processing. Our ``sparse selective strategy'' refers to transmitting only those audio segments the near-sensor model deems as audio-of-interest, effectively filtering out irrelevant data and reducing costly cloud-side computation. What distinguishes our work is the utilization of a near-sensor principle, wherein the sensor selectively collects relevant acoustic data locally, significantly reducing the need for data transmission and processing at the central server. This marks the first attempt to detect audio of interest based on this near-sensor principle, presenting a novel and efficient approach to intelligent sensing in audio data.

\subsection{Hyperdimensional Computing} 
HyperDimensional Computing (HDC), inspired by the brain and modeled with hypervectors~\cite{kanerva2009hyperdimensional}, offers a robust computational paradigm applicable to a wide range of learning problems. This includes areas such as speech recognition~\cite{imani2017voicehd}, genome sequence alignment~\cite{kim2020geniehd}, graph learning~\cite{kang2022relhd, nunes2022graphhd}, clustering~\cite{yun2023hyperdimensional}, and computer vision~\cite{hersche2022constrained, dutta2022hdnn, yun2024spatial}. HDC is known for its high memorization capability, robustness against noise, and model interpretability. It builds upon a well-defined set of operations with random hypervectors, making it extremely resilient to failures and adaptable to diverse computational tasks.

Despite its advantages, the application of HDC to audio detection problems has not been extensively explored. This study aims to leverage the unique properties of HDC for gunshot audio detection, addressing a critical public safety concern. By utilizing HDC's efficient encoding and high-dimensional representation, we aim to develop a model that can accurately detect gunshot sounds in real time, ensuring rapid response and increased safety. This work represents a novel application of HDC, extending its proven benefits in other domains to the field of audio detection and offering a promising solution for urgent, real-world challenges in public safety.

%% file: Chapters/3_Model_Design.tex
In many scenarios, complex machine learning tasks demand heavy models that prove challenging to implement on edge devices. For instance, a recent state-of-the-art Transformer-based audio classification model~\cite{9746312}, demands 80 hours of training on 4 NVIDIA Tesla V100 GPUs, despite its computational resource reduction compared to alternative models. Similarly, many other works also focus on utilizing heavy machine learning models for complex tasks such as using large pre-trained multi-modality model~\cite{wu2022wav2clip}, having large transformer models~\cite{baade2022mae}, etc. Consequently, these contemporary deep learning-based audio detection tasks present a practical challenge when it comes to real-time implementation on edge sensors. Our solution simplifies this by binarizing such tasks, specifically detecting "audio of interest," only essential audio data for complex functions. Unlike conventional models like Recurrent Neural Networks (RNNs), our near-sensor model employs an HDC model with very few CNN layers. This design ensures fast, efficient inference with online learning capability.

Unlike MLPs, HDC does not rely on fully connected layers or activation functions. Instead, it builds class hypervectors directly via bundling and binding operations on encoded features extracted from CNN layers. This approach does not necessitate large parameter sets and backpropagation, enabling rapid adaptation and robust performance even with limited training samples.

\subsection{HDC Basics}
The fundamental representational unit of HDC is called a hyperdimensional vector. A hypervector $\mathcal{H}$ indicates a vector $\mathbb{R}^D$ with high dimensionality $D$. The hyperdimensional vectors are compared to each other by a similarity function $\delta$. Utilizing the similarity measure, HDC can facilitate cognitive tasks such as memorization, classification, clustering, and more. HDC frameworks designed to support these tasks rely on three fundamental HDC operations that directly correspond to brain functionalities: bundling, binding, and permutation.

\begin{enumerate}
    \item \textbf{Bundling}: this operation, denoted as $+$, is typically implemented as element-wise addition. If $\mathcal{H}=\mathcal{H}_1+\mathcal{H}_2$, then both $\mathcal{H}_1$ and $\mathcal{H}_2$ are similar to $\mathcal{H}$. From a cognitive perspective, it can be interpreted as memorization.
    \item \textbf{Binding}: this operation, denoted as $*$, is typically implemented as element-wise multiplication. If $\mathcal{H}=\mathcal{H}_1*\mathcal{H}_2$, then $\mathcal{H}$ is dissimilar to both $\mathcal{H}_1$ and $\mathcal{H}_2$. Binding also has the important property of similarity preservation.
    \item \textbf{Permutation}: this operator, denoted as $\rho$, is typically implemented as a rotation of vector elements.
\end{enumerate}

Using these three operations enables a hyperdimensional learning framework. Classification involves encoding input features into hypervectors and creating class hypervectors by bundling. Retraining involves adjusting class hypervectors by adding hypervectors of correctly predicted samples and subtracting those of misclassified samples, thus refining the class boundaries.

The encoding function often uses cosine and sine transformations, as this mapping preserves similarity between inputs. By projecting data into a high-dimensional space, small differences become more distinguishable, enabling robust recognition and generalization even with limited training data.

\subsection{Bridging HDC to Audio Detection}
While the basics of HDC provide the fundamental building blocks, applying them directly to audio detection involves integrating CNN-based feature extraction with hypervector encoding. The next section shows how we embed HDC operations into an audio sensing framework, transforming raw audio streams into high-dimensional representations and selectively transmitting only data deemed relevant.

\begin{figure*}[t!]
  \centering
  \includegraphics[width=\linewidth]{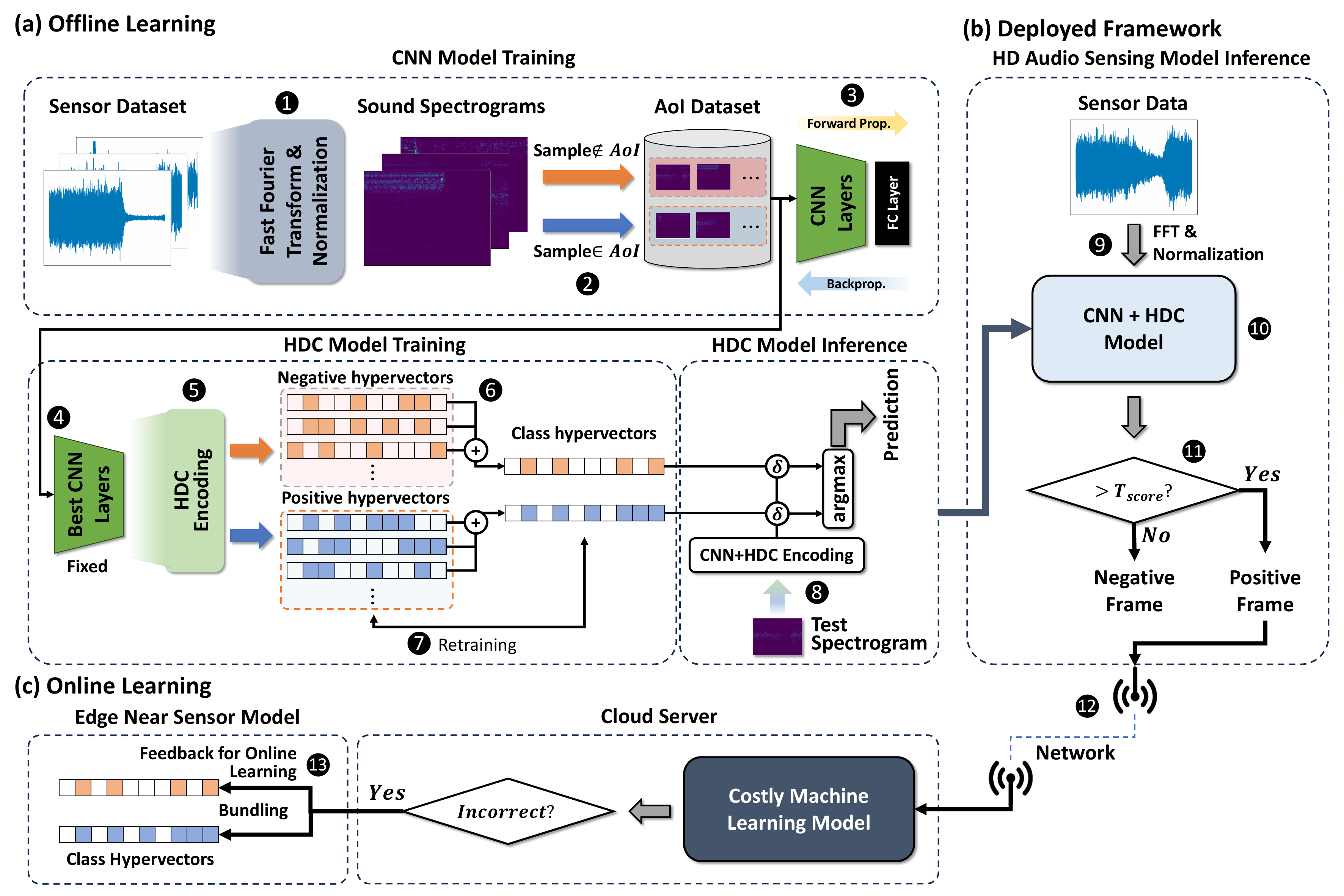}
  \caption{Overview of our audio detection framework for Hyperdimensional Intelligent Sensing. The audio detection training consists of three phases: (a) Offline learning, (b) Offline trained near-sensor model deployment, and (c) Online learning based on a costly machine learning model. After training CNN layers for feature extraction, the HDC encoding transforms extracted features into hypervectors, forming class hypervectors without any traditional MLP layers or activation functions.}
  \label{Fig:model_pipeline}
\end{figure*}

\begin{figure*}
    \centering    \includegraphics[width=0.6\linewidth]{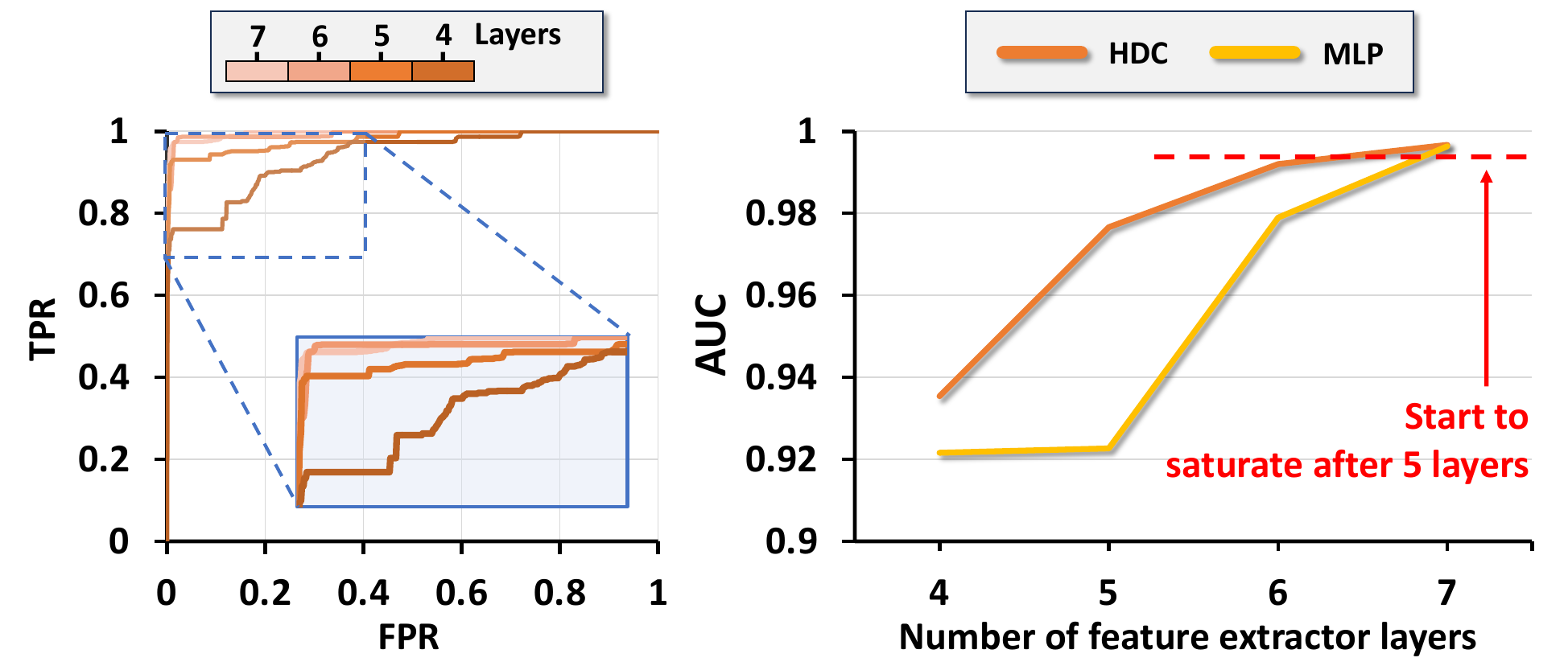}
    \caption{Performance analysis by model size with hyperdimension of $D=10K$. Left: Receiver Operating Characteristic (ROC) curve analysis with varied feature extraction layers. Right: Area Under the Curve (AUC) analysis also with the same range of feature extraction layers.}
    \label{fig:AUCbyModelSize}
\end{figure*}

\begin{figure}
    \centering    \includegraphics[width=1.\linewidth]{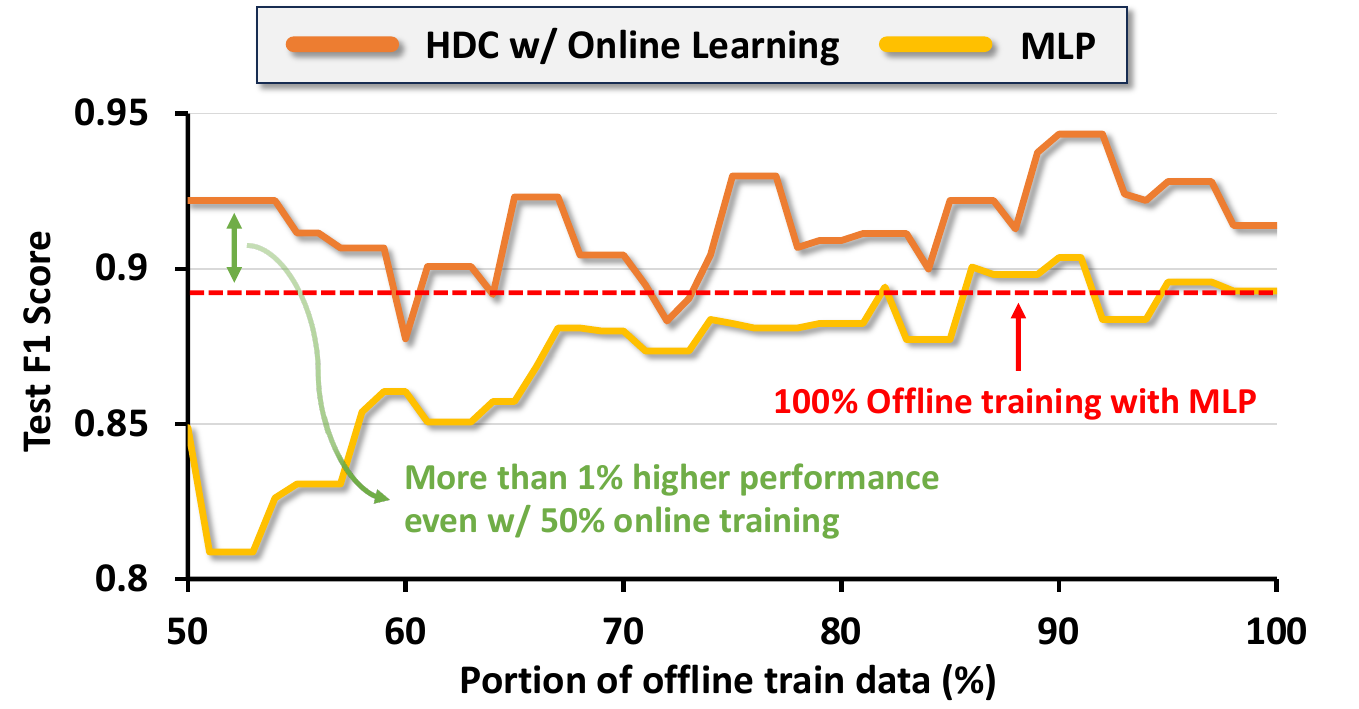}
    \caption{Test F1 score comparison between HDC with online learning and with MLP layer which is hard to support online learning.}
    \label{fig:onlinelearning}
\end{figure}

\begin{figure}
    \centering    \includegraphics[width=1.\linewidth]{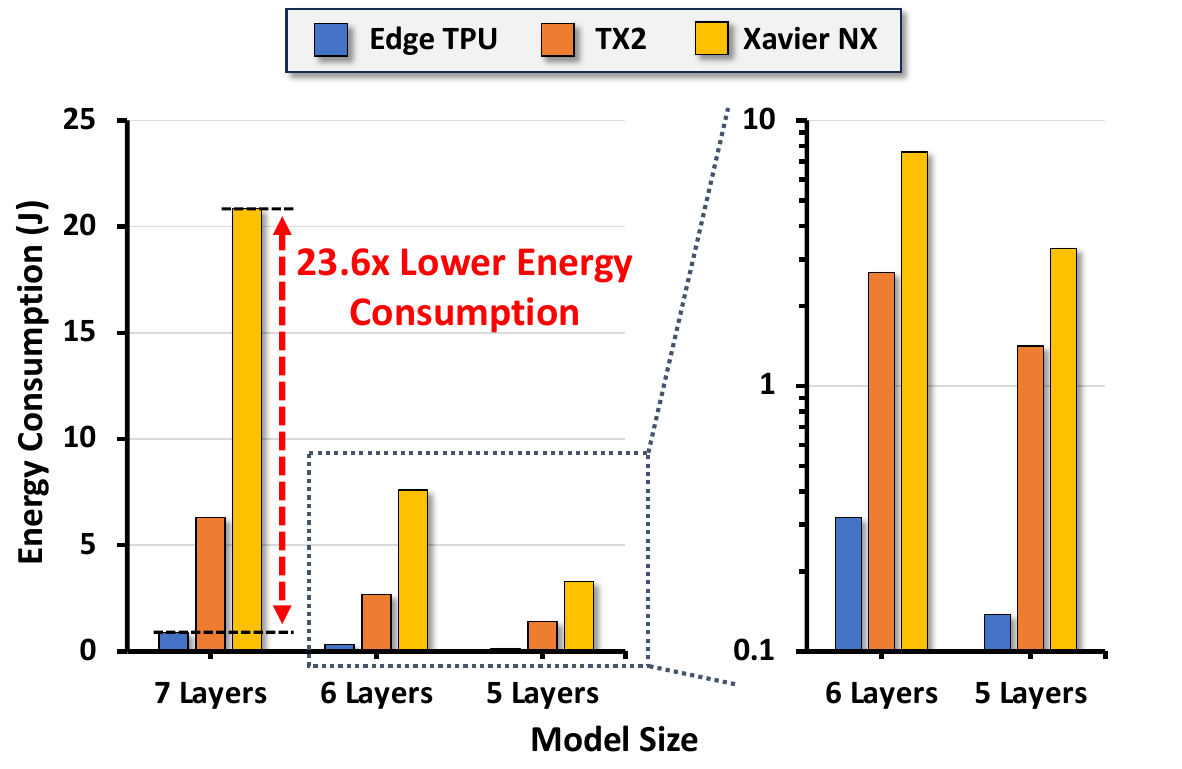}
    \caption{Energy consumption estimation by different near sensor model size}
    \label{fig:energy_consumption}
\end{figure}

\subsection{Audio Intelligent Sensing Framework}

In \autoref{Fig:overview_diagram_ours}, we present an overview of our framework designed to reduce overall system costs related to network communication, expensive machine learning servers, and storage. This is achieved by strategically placing a lightweight AI model in proximity to the microphone sensor, enabling real-time selective transmission of audio data. Leveraging HDC as our lightweight AI model tightly integrated with the sensing circuit, our framework supports online learning, enhancing its adaptability.

To ensure coordination between the lightweight model and the buffer, we maintain a buffer size that matches or exceeds the model's maximum inference latency. The buffer operates as a FIFO queue, and the model processes incoming audio frames in order. Since the model's inference is lightweight and near real-time, it completes classification before the oldest data in the buffer is popped. This synchronization prevents data loss and ensures that the model does not miss any segments it needs to evaluate. In rare high-load scenarios, the buffer size can be increased to handle temporary spikes, ensuring that all data is classified before removal.

Our proposed framework stacks the audio stream in a fixed-size buffer and pops the oldest audio stream data from the buffer if it reaches maximum capacity. During this process, the lightweight AI model determines whether there is audio of interest or not. If the model detects audio of interest, it activates the switch to send out all of the audio data in the buffer through the network communication channel to the costly machine learning server. Adjusting the buffer size not only allows for more contextual data if needed but also helps mitigate false negative detections, particularly useful when audio of interest exhibits time locality characteristics.

\begin{figure*}
    \centering    \includegraphics[width=0.7\linewidth]{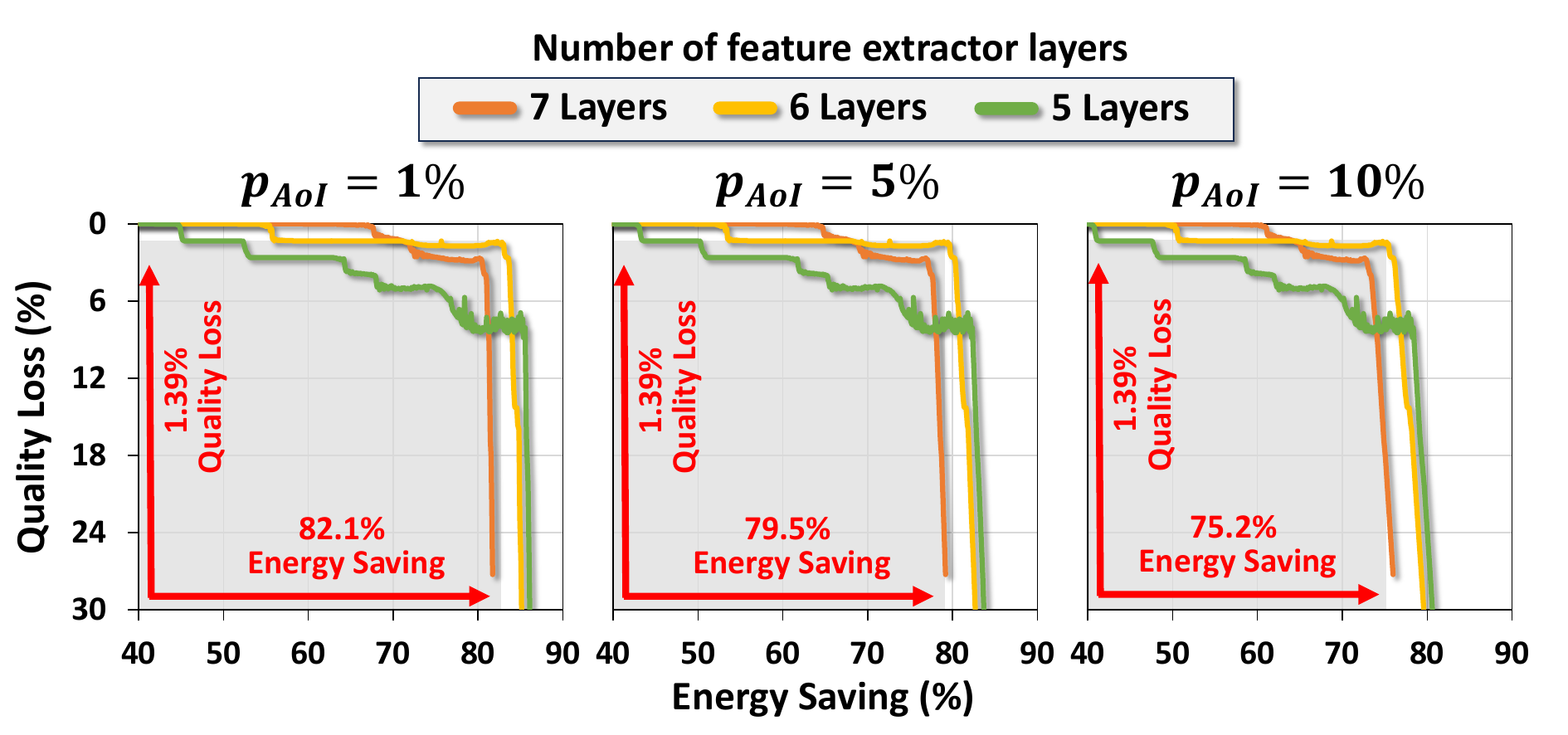}
    \caption{Trade-off relationship between energy saving compared to the conventional method and quality loss.}
    \label{fig:tradeoff}
\end{figure*}

\subsection{Near Audio Sensor Model}

As highlighted earlier, our framework governs data transmission by identifying audio of interest with time locality features. \autoref{Fig:model_pipeline} illustrates the comprehensive pipeline of our near audio sensor model, encompassing three pivotal phases: (a) offline learning, (b) deployed framework, and (c) online learning. In the initial phase, the model undergoes training with an existing audio dataset before transitioning into deployment. Post-training, our model systematically adjusts its weights based on feedback from the resource-intensive machine learning model to sustain optimal performance over time.

In order to deploy the near audio sensor model, we first need to train the model in an offline manner as shown in the \autoref{Fig:model_pipeline}.(a). First, given an audio dataset $D$, we convert them to sound spectrograms using the Fast Fourier Transform (FFT) algorithm with normalization. Now, we generate a labeled dataset by labeling each data according to the Audio of Interest (AoI). For unbalanced scenarios, we use simple random oversampling on the minority AoI samples to ensure that the CNN layers receive sufficient positive samples during training.

After training the CNN layers, we use them as a feature extractor of our HDC model. Using these CNN layers combined with HDC encoding, we generate hypervectors. Negative and positive class hypervectors are formed by bundling their respective sets of hypervectors. To further refine the HDC model, we retrain by incrementally adjusting class hypervectors. This process effectively ``moves'' the class representations closer to correct samples and away from incorrect ones, thereby sharpening decision boundaries without requiring backpropagation or complex layers.

In the deployed framework, the lightweight near-sensor model applies FFT and CNN-based feature extraction on incoming audio. The extracted features are encoded into hypervectors and then compared against class hypervectors. If the similarity score with the positive class hypervector surpasses a threshold $T_{score}$, the data is considered audio of interest and transmitted. We determine $T_{score}$ from ROC curve analysis on a validation set, choosing a threshold that yields an acceptable trade-off between false positives and false negatives.

Finally, online learning is facilitated by the cloud-based heavyweight model. If the cloud model identifies misclassifications from the edge device, it provides feedback hypervectors that are used to update class hypervectors at the edge. This incremental learning allows the near-sensor model to adapt rapidly to evolving data distributions.

\subsection{ASIC Acceleration}
To deploy our model in a resource-constrained edge environment, we employ the Google Edge TPU (Edge-TPU) to accelerate it. We quantize the model into 8-bit integers using the TensorFlow Lite framework. Compared to CPUs and GPUs, ASIC and Edge-TPU approaches dramatically reduce energy consumption due to their specialized architectures. In later sections, we quantitatively show that our approach outperforms conventional embedded CPUs and GPUs by a large margin, confirming that quantization and ASIC integration yield substantial efficiency gains.

For clarity, in a practical scenario, the microphone feeds raw audio to the near-sensor ASIC, which performs FFT, CNN feature extraction, and HDC classification. Only detected AoI data is sent to the cloud for heavyweight processing. We have not built a custom hardware testbed but rely on known power profiles and simulation results to estimate energy savings.

%% file: Chapters/4_Experiments.tex
\begin{figure*}[t!]
    \centering    \includegraphics[width=0.75\linewidth]{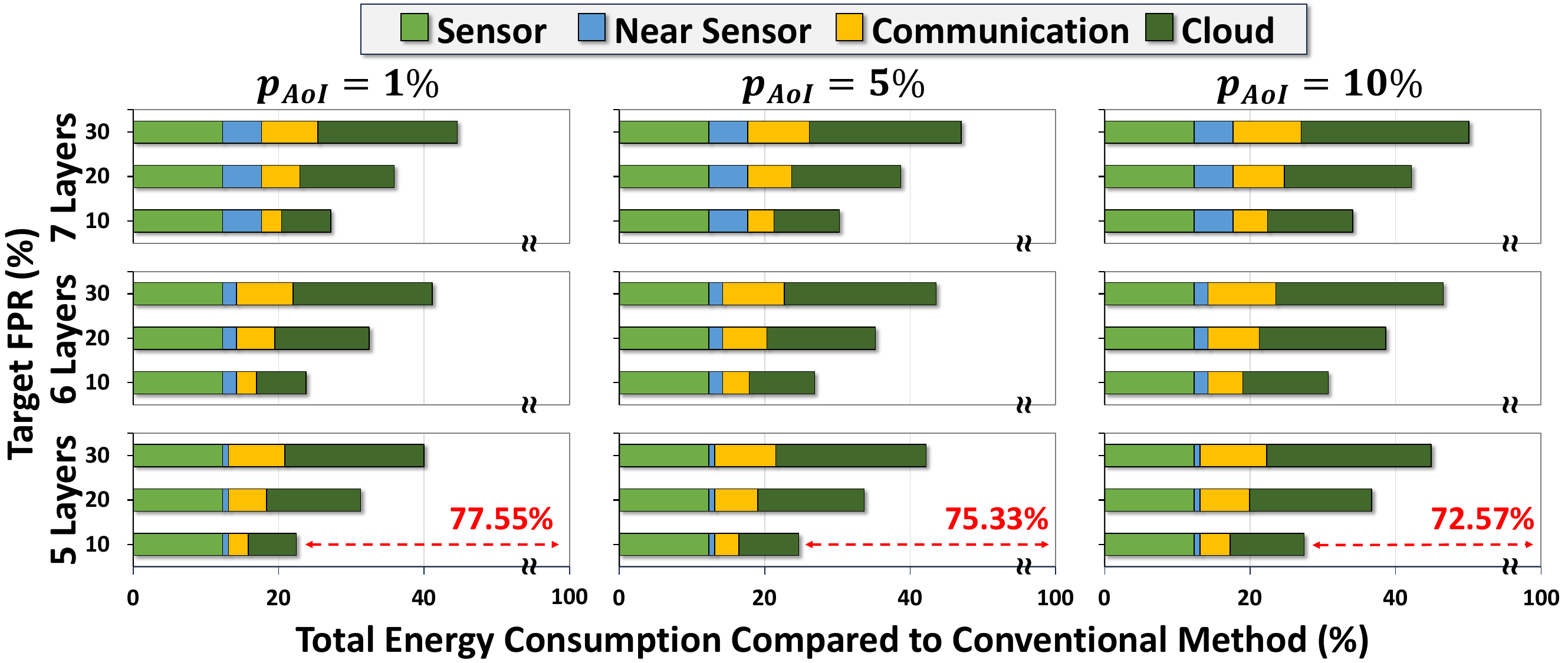}
    \caption{Our framework's energy breakdown in edge client and cloud server communication scenario when normalizing energy consumption by the total energy consumption of the conventional method. Unlike the conventional method, our model reduces cloud processing because fewer irrelevant segments are transmitted. In scenarios with smaller near-sensor models, the reduction in communication and cloud energy might be limited if the false positive rate remains high, but our largest model configurations achieve balanced reductions across edge, communication, and cloud.}
    \label{fig:ours_energybreakdown}
\end{figure*}

\begin{figure*}[t!]
    \centering    \includegraphics[width=0.75\linewidth]{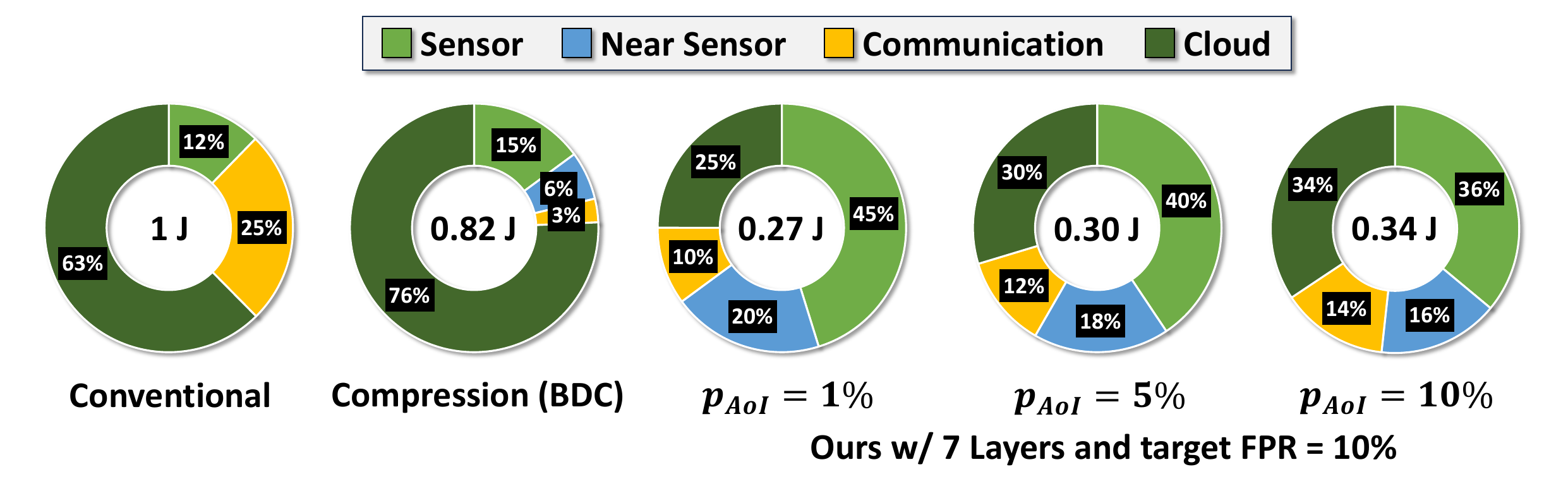}
    \caption{Distribution of energy consumption for the conventional method, compressive near sensor method, and ours in different AoI probabilities. Total energy consumptions are presented in the centers normalized by the conventional method's total energy consumption. The conventional method transmits all data to the cloud, resulting in high communication and cloud-side processing costs. Our approach significantly reduces total energy by filtering data at the edge.}
    \label{fig:pie_breakdown}
\end{figure*}

\subsection{Experimental Setup}
The proposed framework has been executed with a software framework and a hardware accelerator. Our software framework is implemented using a combination of PyTorch and NumPy that supports CNN layers, HDC encoding, and classification. We study the effectiveness of our technique over the UrbanSound8K dataset~\cite{salamon2014dataset}, which is a public audio dataset for urban sound classification applications. The dataset contains 10 classes such as car horn, dog bark, etc. Among these classes, we chose to use the gunshot class as our audio of interest.

\subsection{Evaluation of The Proposed Near-sensor Model}

In the evaluation of our near-sensor model, we focused on a scenario where our objective is to permit a specific level of false positives while maximizing the true positive rate (TPR). Employing the ROC curve evaluation allows us to analyze this scenario and identify model configurations that achieve the highest TPR at a desired false positive rate (FPR).

Compared to an MLP, HDC achieves better training results even with a limited portion of training data because HDC relies on high-dimensional random projections that preserve similarity and enable robust classification with fewer samples. MLPs typically need more data to effectively adjust their parameters via gradient-based training. In contrast, HDC can immediately construct meaningful class hypervectors from a small number of examples, providing inherent robustness and reducing the data requirement.

Our initial focus was on evaluating the impact of the feature extractor size, comprised of CNN layers, on achieving a sharp ROC curve. Even with slightly more than 5 layers of CNN, the model achieves an AUC exceeding 0.99, with saturation evident at approximately the maximum AUC. The HDC model displays superior learning ability over the MLP reference, evidenced by its enhanced AUC performance.

\autoref{fig:onlinelearning} demonstrates the online learning capability of the HDC model. The HDC model can easily update its class hypervectors as new data arrives. In contrast, the MLP model requires additional mechanisms for reliable online adaptation. HDC with online learning surpasses the offline-trained MLP model in terms of F1 score on the test set, showing HDC’s inherent advantage in incremental learning scenarios.

\subsection{Evaluation of Energy Efficiency using ASIC Design}

To optimize the overall system's energy efficiency, the development of an energy-conscious near-sensor model is imperative. We conducted a comparative energy consumption analysis, contrasting our ASIC implementation on the Edge TPU with two widely utilized boards equipped with edge-specific CPUs and GPUs. The Edge TPU exhibits approximately 23.6 times greater energy efficiency than some counterparts.

In \autoref{fig:ours_energybreakdown} and \autoref{fig:pie_breakdown}, the baseline or ``conventional'' method means sending all raw data to the cloud for processing without any near-sensor filtering. By comparing our method’s bars to this baseline, it becomes clear that while smaller models may not significantly reduce communication or cloud energy (due to fewer AoI detections), larger or more tuned models do. The figure shows normalization by the conventional method’s total energy, making direct comparisons straightforward.

As shown in \autoref{fig:ours_energybreakdown} and \autoref{fig:pie_breakdown}, our framework demonstrates significant energy savings over conventional and compression-based approaches. Even with small models and carefully chosen FPR thresholds, we can achieve over 77\% energy reduction. By adjusting $T_{score}$ to allow slightly higher FPR, we can further improve TPR, capturing more AoI samples and achieving up to 82.1\% energy savings at the cost of only 1.39\% quality loss, as shown in \autoref{fig:tradeoff}.

%% file: Chapters/5_Conclusions.tex
We present the Hyperdimensional Intelligent Sensing Framework, designed for efficient real-time audio processing at the extreme edge. This innovative framework aims to significantly reduce overall system costs, including energy and storage consumption, by incorporating a near-sensor model focused on detecting essential audio data. This selectively transmitted data, crucial for the target task, is efficiently processed on a high-performance cloud server. Our evaluation demonstrates the framework's extreme efficiency, achieving up to 82.1\% energy savings with only 1.39\% quality loss. The near-sensor model's efficacy is highlighted by its sharp ROC curve, indicating robust performance. Notably, the ASIC implementation of the near-sensor model exhibits exceptional energy efficiency, confirming the framework's ability to substantially decrease total system energy consumption while maintaining a balanced trade-off between energy conservation and quality preservation.

In future work, we plan to explore more sophisticated threshold selection strategies and consider integrating advanced HDC encodings for even greater robustness. Additionally, constructing a physical prototype and conducting on-site experiments will help validate our simulation-based findings and further refine the system for practical deployments.